\documentclass[12pt]{article}

\begin{document}

\begin{titlepage}

\begin{flushright}
IUHET-468\\
\end{flushright}
\vskip 2.5cm

\begin{center}
{\Large \bf Compton Scattering in the Presence of \\
Lorentz and CPT Violation}
\end{center}

\vspace{1ex}

\begin{center}
{\large B. Altschul\footnote{{\tt baltschu@indiana.edu}}}

\vspace{5mm}
{\sl Department of Physics} \\
{\sl Indiana University} \\
{\sl Bloomington, IN 47405 USA} \\

\end{center}

\vspace{2.5ex}

\medskip

\centerline {\bf Abstract}

\bigskip

We examine the process of Compton scattering, in the presence of a Lorentz-
and CPT-violating modification to the structure of the electron. We calculate
the complete tree-level contribution to the cross section; our
result is valid to all orders in the Lorentz-violating parameter. We find a
cross section that differs qualitatively from the Klein-Nishina result at
small frequencies, and we also encounter a previously undescribed complication
that will arise in the calculation of many Lorentz-violation cross sections:
The Lorentz violation breaks the spin degeneracy of the external states, so we
cannot use a closure relation to calculate the unpolarized cross section.

\bigskip

\end{titlepage}

\newpage

Recently, there has been a great deal of interest in the possibility of there
existing small CPT- and
Lorentz-violating corrections to the standard
model~\cite{ref-kost1,ref-kost2,ref-kost3,ref-kost4}. Such small
corrections might
arise from larger violations of Lorentz symmetry occurring at the Planck scale.
The most general possible Lorentz-violating effective field
theory has been described
in detail, and its renormalizability has been studied. These results open the
way for a wide variety of experiments that could test for the existence of
Lorentz violation.

There are already many
experimental constraints on Lorentz-violating corrections to the standard
model. The tests have included studies of matter-antimatter asymmetries for
trapped charged particles~\cite{ref-bluhm1,ref-bluhm2,ref-gabirelse,
ref-dehmelt1} and bound state systems~\cite{ref-bluhm3,ref-phillips},
frequency standard comparisons~\cite{ref-berglund,ref-kost6,ref-bear},
measurements of neutral meson oscillations~\cite{ref-kost7,ref-hsiung,ref-abe},
polarization measurements on the light from distant galaxies~\cite{ref-carroll1,
ref-carroll2}, and many others. However, although there have been a number of
kinematical analyses of the astrophysical consequences of
Lorentz violation in particle
scattering~\cite{ref-kif,ref-aloisio,ref-amelino,ref-jacob,ref-konopka,ref-hey},
there has as yet been very little
investigation into the possible effects of Lorentz-violating dynamics in
laboratory scattering
experiments~\cite{ref-arfaei,ref-kost5,ref-kersting}. In this paper, we shall
examine some of those effects.

We shall examine the process of Compton scattering, in the presence of a
particular Lorentz- and CPT-violating modification of the electron sector.
The study of Compton scattering has historically been very important to the
development of quantum mechanics and quantum field theory~\cite{ref-compton,
ref-debye,ref-klein} and in the future might provide an important test of
Lorentz violation.

The calculation of scattering cross sections in a Lorentz-violating theory
involves a number of subtleties that are not present in the standard,
Lorentz-invariant case. Different reference frames are no longer necessarily
equivalent, and the correct definition of the particle flux becomes potentially
ambiguous. However, with appropriate care, meaningful cross sections can be
found, and a general theory for their calculation is given
in~\cite{ref-kost5}.
Our analysis will also reveal an additional complication, not
discussed in~\cite{ref-kost5}, that may arise in Lorentz-violating scattering
processes. The various spin states of the scattered particles may have
differing energies, and this can affect the velocity and
phase space factors that appear in
the cross section. As a result, it may become
impossible to calculate an unpolarized
cross section by the usual means.

The Lagrange density for our theory is
\begin{equation}
\label{eq-L}
{\cal L}=-\frac{1}{4}F^{\mu\nu}F_{\mu\nu}+\bar{\psi}(i\!\!\not\!\partial-m-
e\!\!\not\!\!A\,-\!\not\!b\gamma_{5})\psi.
\end{equation}
The action includes only the single Lorentz-violating coefficient $b^{\mu}$.
This is the simplest perturbatively nontrivial form of Lorentz violation that
can exist in the electron sector. The domain of validity of this Lagrange
density extends all the way up the Planck scale~\cite{ref-kost3}.

Considering a theory with only a $b$ term would
not be reasonable for calculations beyond tree level; other Lorentz-violating
terms would be radiatively generated at one-loop order~\cite{ref-kost4}. We
shall therefore consider only tree-level effects. However, although we shall
only be working to leading order in the electromagnetic coupling $e^{2}$, our
results will be correct to all orders in $b$.

In general, the spacetime direction of $b$ is arbitrary. However, we
shall choose $b$ to be purely timelike, $b^{\mu}=(B,\vec{0}\,)$, in the
laboratory
frame. It is a common practice to suppose that any Lorentz-violating
coefficients have vanishing spatial components; this practice arises from the
observation that the universe shows a very high degree of isotropy in the rest
frame of the cosmic microwave background. In this case, considering only a
timelike $b$ will also substantially simplify our $b$-exact analysis of the
theory.

Since our Lorentz-violating Lagrange density (\ref{eq-L})
involves no changes to the electrons' kinetic term, and there are no additional
time derivatives not present in the Lorentz-invariant theory, the electrons may
be quantized without any changes to the spinor
representation~\cite{ref-kost3,ref-kost5}. The exact electron propagator may be
read off directly from the Lagrange density; it is
\begin{equation}
S(l)=\frac{i}{\!\not l-m\,-\!\not\!b\gamma_{5}}.
\end{equation}
We may rationalize this expression and obtain~\cite{ref-victoria1,ref-chung1}
\begin{equation}
\label{eq-propagator}
S(l)=i\frac{(\!\not l+m\,-\!\not\!b\gamma_{5})(l^{2}-m^{2}-b^{2}+[\!\not l,
\!\not\!b\,]\gamma_{5})}{(l^{2}-m^{2}-b^{2})^{2}+4[l^{2}b^{2}-(l\cdot b)^{2}]}.
\end{equation}
This modification of the propagator represents one of the ways in which the
presence of $b$ will affect the theory.

However, before we can investigate how the modified propagator $S(l)$
affects the dynamics of the scattering, we must examine the effects of the
Lorentz violation on the kinematics.
The coefficient $b$ will affect the structure of the
theory's asymptotic states. The photon states are, of course, unaffected, but
the incoming and outgoing spinors will be significantly modified. We must solve
the free momentum-space Dirac equation, with the $b$ term included, to
determine the propagation modes of the electrons. Since the matrix $\gamma_{5}$
features prominently in the theory, it is natural
to use the Weyl chiral
representation for the Dirac matrices:
\begin{equation}
\label{eq-gamma}
\gamma^{0}=\left[
\begin{array}{cc}
0 & 1 \\
1 & 0
\end{array}\right],
\, \gamma^{i}=\left[
\begin{array}{cc}
0 & \sigma^{i} \\
-\sigma^{i} & 0
\end{array}\right],
\, \gamma_{5}=\left[
\begin{array}{cc}
-1 & 0 \\
0 & 1
\end{array}\right].
\end{equation}
For an electron mode with energy $E$ and three-momentum $\vec{p}=p_{3}\hat{z}$,
with $p_{3}\geq 0$,
the Dirac equation may be reduced to
\begin{equation}
\left[
\begin{array}{cc}
E+B+p_{3}\sigma^{3} & -m \\
-m & E-B-p_{3}\sigma^{3}
\end{array}
\right]u(p)=0.
\end{equation}
If the electron has spin $\frac{s}{2}$ along the $z$-axis, then we may replace
$\sigma^{3}\rightarrow s$. The eigenvalue condition for $E$ then becomes
\begin{equation}
\label{eq-E}
E^{2}=m^{2}+(sp_{3}+B)^{2}=m^{2}+(s|\vec{p}\,|+B)^{2},
\end{equation}
and the spinor is
\begin{equation}
u^{s}(p)=\left[
\begin{array}{c}
\sqrt{\sqrt{m^{2}+(sp_{3}+B)^{2}}-(sp_{3}+B)}\,\xi_{s} \\
\sqrt{\sqrt{m^{2}+(sp_{3}+B)^{2}}+(sp_{3}+B)}\,\xi_{s}
\end{array}
\right],
\end{equation}
where the $\xi_{s}$ are basis spinors quantized in the $z$-direction. This
solution may easily
be generalized to describe electrons with arbitrary three-momentum,
so long as the spin is quantized along the direction of the motion. Our spinors
satisfy the conventional normalization
conditions $\bar{u}^{s'}(p)u^{s}(p)=2m\delta^{ss'}$ and
$u^{s'\dag}(p)u^{s}(p)=2E(p)\delta^{ss'}$.
Note that even though there is no breaking of rotation invariance, the energy
depends upon the spin direction, through the helicity $s$.

Just as in the Lorentz-invariant case, a great deal can be learned
about the scattering simply from an analysis of the energy-momentum relation
(\ref{eq-E}).
Let us consider an experiment in which the initial electron has vanishing
three-momentum---$p=(E,\vec{0}\,)=(\sqrt{m^{2}+B^{2}},\vec{0}\,)$; then the
incoming spinor is
\begin{equation}
u_{i}^{s}(p)=\left[
\begin{array}{c}
\sqrt{\sqrt{m^{2}+B^{2}}-B}\,\xi_{s} \\
\sqrt{\sqrt{m^{2}+B^{2}}+B}\,\xi_{s}
\end{array}
\right].
\end{equation}
However, even though the electron's three-momentum vanishes, it is not really
stationary; because of the Lorentz-violation,
the group velocity for a wave packet centered around $\vec{p}\,=\vec{0}$ is
nonvanishing, and we must account for this velocity in the definition of the
electron flux. In general, Lorentz-violating effects could also cause the
velocity and three-momentum of the electron to point in different directions,
but in this case, the two quantities are always collinear.

The almost stationary electron is struck by a photon with momentum
$k^{\mu}=(\omega,\omega\hat{z})$.
(This is a reasonable setup for a low- or medium-energy experiment.)
The photon is scattered through an angle $\theta$ and has
outgoing momentum $k'^{\mu}=(\omega',\omega'\sin\theta,0,\omega'\cos\theta)$.
The scattered electron has three-momentum $\vec{p}\,'$, and the corresponding
adjoint spinor is
\begin{equation}
\bar{u}_{f}^{s'}(p')=\left[
\begin{array}{c}
\sqrt{\sqrt{m^{2}+(s'|\vec{p}\,'|+B)^{2}}+(s'|\vec{p}\,'|+B)}\,\xi'^{*}_{s'} \\
\sqrt{\sqrt{m^{2}+(s'|\vec{p}\,'|+B)^{2}}-(s'|\vec{p}\,'|+B)}\,\xi'^{*}_{s'}
\end{array}
\right]^{T},
\end{equation}
where $\frac{s'}{2}$ is the spin, quantized along the $\vec{p}\,'$-direction.
For the sake of brevity, we define $C=s'|\vec{p}\,'|+B$.
The energy of the scattered electron is $E'=\sqrt{m^{2}+C^{2}}$.

We may now derive a generalization of Compton's
wavelength shift relation, $1-\cos\theta=m(1/\omega'-1/\omega)$.
From three-momentum conservation, we have that $\omega'\sin\theta=|\vec{p}\,'|
\sin\Theta$ and $\omega-\omega'\cos\theta=|\vec{p}\,'|\cos\Theta$, where
$\Theta$ is the angle through which the electron is scattered. (More precisely,
$\Theta$ is the angle describing the orientation of the momentum. If
$|\vec{p}\,'|$ is small enough
that $|\vec{p}\,'|+sB<0$, then the three-momentum and the
group velocity are oriented in opposite directions, so the correct scattering
angle is $\pi-\Theta$.)
Taken together, the equations for $\Theta$
give us
\begin{equation}
\label{eq-Theta}
\Theta=\tan^{-1}\frac{\sin\theta}{\frac{\omega}{\omega'}-\cos\theta}
\end{equation}
and
\begin{equation}
\label{eq-absp'}
|\vec{p}\,'|^{2}=\omega^{2}+(\omega')^{2}-2\omega\omega'\cos\theta;
\end{equation}
these equations  do not depend upon $B$. Using (\ref{eq-absp'}), the
energy conservation condition becomes
\begin{equation}
\label{eq-Econs}
\omega+\sqrt{m^{2}+B^{2}}=\omega'+\sqrt{m^{2}+\left[s'\sqrt{\omega^{2}+(\omega')
^{2}-2\omega\omega'\cos\theta}+B\right]^{2}}.
\end{equation}
By repeatedly squaring the equation (\ref{eq-Econs}), we can arrive at a
quadratic equation for $(1-\cos\theta)$:
\begin{equation}
\label{eq-cosquadratic}
\omega^{2}(\omega')^{2}(1-\cos\theta)^{2}-2\omega\omega'\left[(\omega-\omega')
\sqrt{m^{2}+B^{2}}+B^{2}\right](1-\cos\theta)+m^{2}(\omega-\omega')^{2}=0.
\end{equation}
Note that the $s'$-dependence of the energy has vanished from this expression.

In the Lorentz-invariant case, $B=0$, equation (\ref{eq-cosquadratic}) is a
perfect square, with only one solution for $(1-\cos\theta)$. For $B\neq0$,
(\ref{eq-cosquadratic}) has two solutions. The correct one may be identified
by noting that forward scattering ($\theta=0$) must correspond to $\omega=
\omega'$. We then see that
\begin{equation}
\label{eq-cos}
1-\cos\theta=\frac{1}{\omega\omega'}\left[(\omega-\omega')\sqrt{m^{2}+B^{2}}+
B^{2}-|B|\sqrt{(\omega-\omega')^{2}+B^{2}+2(\omega-\omega')\sqrt{m^{2}+B^{2}}}
\right].
\end{equation}
This relation represents the modification of the Compton effect caused by the
presence of the Lorentz-violating coefficient $b$.
If we then expand (\ref{eq-cos}) to first order in $B$, we find
\begin{equation}
\label{eq-cosapprox}
1-\cos\theta\approx m\left(\frac{1}{\omega'}-\frac{1}{\omega}\right)\left[1-
\frac{|B|}{m}\sqrt{1+\frac{2m}{(\omega-\omega')}}\right].
\end{equation}
Alternatively, to
obtain the ${\cal O}(B)$ correction to the Compton wavelength shift, we may
replace $\omega-\omega'$ in (\ref{eq-cosapprox}) with the $B=0$ expression
\begin{equation}
\omega-\omega'=\omega\frac{1-\cos\theta}{\frac{m}{\omega}+(1-\cos\theta)}.
\end{equation}
This gives $1/\omega'-1/\omega$ as a function of $|B|$, $\omega$, and $(1-\cos
\theta)$.
So the relations (\ref{eq-Theta}), (\ref{eq-absp'}), and (\ref{eq-cos}) are
sufficient to express all the kinematically constrained variables in the problem
in terms of a single quantity---either
the scattered photon's energy $\omega'$ or the
scattering angle $\theta$.

To complete our discussion of the kinematics, we must determine the flux
normalization and phase space factors that appear in the differential cross
section. These factors account for the properties of the initial and final
states, respectively. The flux normalization factor in the cross section is
$1/F$, where
\begin{equation}
F=N_{\gamma}N_{e}|\vec{v}_{\gamma}-\vec{v}_{e}|.
\end{equation}
$N_{\gamma}$ and $N_{e}$ are the photon and electron beam densities, while
$\vec{v}_{\gamma}$ and $\vec{v}_{e}$ are the corresponding velocities in the
laboratory frame. All the particles obey conventional normalization
conditions, so $N_{\gamma}=2\omega$ and $N_{e}=2E=2\sqrt{m^{2}+B^{2}}$.
The photon velocity is clearly $\vec{v}_{\gamma}=\hat{z}$, and the group
velocity for the elctron wave packet is
\begin{equation}
\label{eq-ve}
\vec{v}_{e}=\frac{sB}{E}\hat{z}.
\end{equation}

The impact of the outgoing states is more subtle. The phase space integral
is~\cite{ref-kost5}
\begin{eqnarray}
\int d\Pi & = &\int\frac{d^{3}k'}{(2\pi)^{3}}\frac{1}{2\omega'}
\frac{d^{3}p'}{(2\pi)^{3}}\frac{1}{E'}(2\pi)^{4}\delta^{4}(k'+p'-k-p) \\
& = & \frac{1}{16\pi^{2}}\int(\omega')^{2}\,d\omega'\,d\Omega
\frac{1}{\omega'E'}\delta(\omega'+E'-\omega-E).
\end{eqnarray}
We use the
$\delta$-function to perform the $\omega'$ integration; this gives us a
factor of
\begin{equation}
\frac{1}{\left|\frac{\partial}{\partial\omega'}(\omega'+E'-\omega-E)\right|}=
\frac{1}{1+\left(1+\frac{s'B}{|\vec{p}\,'|}\right)\frac{\omega'-\omega\cos
\theta}{E'}}.
\end{equation}
Therefore, the phase space factor is given by
\begin{equation}
\label{eq-phasesp}
\int d\Pi = \frac{1}{16\pi^{2}}\int d\Omega\frac{\omega'}{E'+\left(
1+\frac{s'B}{|\vec{p}\,'|}\right)(\omega'-\omega\cos\theta)}.
\end{equation}

(\ref{eq-ve}) and (\ref{eq-phasesp}) depend explicitly upon $s$ and
$s'$, so the impact velocity and the available phase space
are not independent of the electron's spin. This is an important
observation, because it affects the way in which we must calculate the cross
section. In high-energy physics experiments, one frequently measures only
unpolarized cross sections. Moreover, the unpolarized formulas are often
especially simple
in form and easy to obtain, thanks to Casimir's trick of using the closure
relation for the Dirac spinors to perform the spin sum. However, in order to use
this trick, the velocity and phase space factors must be independent of the
incoming and outgoing polarizations.
In the situation we are considering, Lorentz violation has broken
the spin degeneracy of the electron states' energies. We therefore cannot use
Casimir's method to sum over the spin states. Instead, we
shall calculate the cross section using a basis of explicit polarization states
for all the incoming and outgoing particles.

This completes our discussion of the Compton scattering kinematics, and
we now turn our attention to the details of the dynamics.
The scattering matrix element ${\cal M}$ is given by
\begin{eqnarray}
i{\cal M} & = & \bar{u}_{f}^{s'}(p')\left(-ie\gamma^{\mu}\right)\epsilon'^{*}
_{\mu}(k')S(p+k)\left(-ie\gamma^{\nu}\right)\epsilon_{\nu}(k)u_{i}^{s}(p)
\nonumber\\
\label{eq-iM}
& & +\,\bar{u}_{f}^{s'}(p')\left(-ie\gamma^{\nu}\right)\epsilon
_{\nu}(k)S(p-k')\left(-ie\gamma^{\mu}\right)\epsilon'^{*}_{\mu}(k')u_{i}^{s}(p),
\end{eqnarray}
where $\epsilon(k)$ and $\epsilon'(k')$ are the polarization vectors for the
external photons. Although we have restricted the electron's spin to be
quantized along its direction of motion, we may allow the photon polarization
basis to remain arbitrary, since there is no tree-level Lorentz-violation in the
electromagnetic sector.

The propagators appearing in (\ref{eq-iM}) have arguments of the form
$l^{\mu}=(\sqrt{m^{2}+B^{2}}+\Omega,\Omega\hat{u})$, where $\Omega=\omega$ or
$-\omega'$, and $\hat{u}$ is a unit three-vector. So the denominator of
(\ref{eq-propagator}) reduces to the extremely simple form
\begin{equation}
(l^{2}-m^{2}-b^{2})^{2}+4[l^{2}b^{2}-(l\cdot b)^{2}]=4\Omega^{2}m^{2}.
\end{equation}
Using our explicit representation (\ref{eq-gamma})
of the gamma matrices, the entire propagator then becomes
\begin{eqnarray}
S(l)& = & \frac{i}{2\Omega m^{2}}\left\{\left[
\begin{array}{cc}
mE & \left(\Omega+E\right)\left(E-B\right) \\
\left(\Omega+E\right)\left(E+B\right) &
mE
\end{array}
\right]\right. \nonumber\\
\label{eq-prop2}
& & +\left.\left[
\begin{array}{cc}
mB & \left(B-\Omega\right)\left(E-B\right) \\
\left(B+\Omega\right)\left(E+B\right) & mB
\end{array}
\right]\sigma_{\hat{u}}\right\},
\end{eqnarray}
where $\sigma_{\hat{u}}=\vec{\sigma}\cdot\hat{u}$ is the Pauli spin matrix
corresponding to the direction $\hat{u}$.

We now need to evaluate $u_{f}^{s'}\gamma^{\alpha}S(l)\gamma^{\beta}u_{i}^{s}$.
The products of Pauli matrices that will arise may be simplified
by noting that,
since the polarization vectors $\epsilon(k)$ and $\epsilon'(k')$ are purely
spacelike, $\alpha$ and $\beta$ will take on only spacelike values.
Setting $\alpha=j$ and $\beta=k$, we see
that we need only evaluate the combinations
\begin{eqnarray}
\label{eq-2sigma}
\sigma^{j}\sigma^{k} & = & \delta^{jk}+i\epsilon^{jkl}\sigma^{l} \\
\sigma^{j}\sigma^{l}\sigma^{k} & = & \delta^{jl}\sigma^{k}+\delta^{kl}\sigma^{k}
-\delta^{jk}\sigma^{l}-i\epsilon^{jkl}
\end{eqnarray}
These expressions are to be contracted with $\vec{\epsilon}\,(k)$,
$\vec{\epsilon}\,'^{*}(k')$, and $\hat{u}$.

Ultimately, there are four terms in the matrix element,
since there are two diagrams in (\ref{eq-iM}) and two
terms in the propagator (\ref{eq-prop2}). We shall evaluate each term
individually. We begin with the contribution ${\cal M}_{1}$
coming from the first terms in both
(\ref{eq-iM}) and (\ref{eq-prop2}). If we contract (\ref{eq-2sigma}) with the
external vectors and use the identity
\begin{equation}
m=\sqrt{\sqrt{m^{2}+B^{2}}+B}\sqrt{\sqrt{m^{2}+B^{2}}-B},
\end{equation}
we find the final expression
\begin{eqnarray}
{\cal M}_{1} & = & \frac{-e^{2}}{2\omega m^{2}}\left\{\sqrt{\sqrt{m^{2}+C^{2}}+
C}\sqrt{\sqrt{m^{2}+B^{2}}+B}\left[(\omega+2B)\sqrt{m^{2}+B^{2}}+\omega B\right]
\right.\nonumber\\
& & +\left.\sqrt{\sqrt{m^{2}+C^{2}}-C}
\sqrt{\sqrt{m^{2}+B^{2}}-B}\left[(\omega-2B)\sqrt{m^{2}+B^{2}}-\omega B\right]
\right\} \nonumber\\
& & \times\left[\left(\vec{\epsilon}\,'^{*}\cdot\vec{\epsilon}\,\right)
\left(\xi'^{\dag}
_{s'}\xi_{s}\right)+i\left(\vec{\epsilon}\,'^{*}\times\vec{\epsilon}\,\right)
\cdot\left(\xi'^{\dag}_{s'}\vec{\sigma}\,\xi_{s}\right)\right].
\end{eqnarray}
The second contribution from the same Feynman diagram is similar:
\begin{eqnarray}
{\cal M}_{2} & = & \frac{-e^{2}}{2\omega m^{2}}\left\{\sqrt{\sqrt{m^{2}+C^{2}}+
C}\sqrt{\sqrt{m^{2}+B^{2}}+B}\left[(\omega+2B)B +\omega\sqrt{m^{2}+B^{2}}\right]
\right.\nonumber\\
& & +\left.\sqrt{\sqrt{m^{2}+C^{2}}-C}
\sqrt{\sqrt{m^{2}+B^{2}}-B}\left[(\omega-2B)B-\omega\sqrt{m^{2}+B^{2}}\right]
\right\} \nonumber\\
& & \times\left[\left(\vec{\epsilon}\,'^{*}\right)_{3}\vec{\epsilon}\cdot
\left(\xi'^{\dag}_{s'}\vec{\sigma}\,\xi_{s}\right)-
i\left(\vec{\epsilon}\,'^{*}
\times\vec{\epsilon}\,\right)_{3}\left(\xi'^{\dag}_{s'}\xi_{s}\right)-
\left(\vec{\epsilon}\,'^{*}\cdot\vec{\epsilon}\,\right)
\left(\xi'^{\dag}_{s'}\sigma^{3}\xi_{s}\right)
\right].
\end{eqnarray}
The other two contributions to ${\cal M}$ are also similar in form. They differ
from ${\cal M}_{1}$ and ${\cal M}_{2}$ in
three ways: We must make the replacement $\omega\rightarrow-\omega'$, reverse
the signs of the cross product terms (because the order of $\gamma^{\mu}$ and
$\gamma^{\nu}$ has been switched), and change $\hat{u}$ from $\hat{z}$ to
$\hat{k}'=(\sin\theta,0,\cos\theta)$. The resulting
contributions are
\begin{eqnarray}
{\cal M}_{3} & = & \frac{-e^{2}}{2\omega' m^{2}}\left\{\sqrt{\sqrt{m^{2}+C^{2}}+
C}
\sqrt{\sqrt{m^{2}+B^{2}}+B}\left[(\omega'-2B)\sqrt{m^{2}+B^{2}}+\omega' B\right]
\right.\nonumber\\
& & +\left.\sqrt{\sqrt{m^{2}+C^{2}}-C}
\sqrt{\sqrt{m^{2}+B^{2}}-B}\left[(\omega'+2B)\sqrt{m^{2}+B^{2}}-\omega' B
\right]\right\} \nonumber\\
& & \times\left[\left(\vec{\epsilon}\,'^{*}\cdot\vec{\epsilon}\,\right)
\left(\xi'^{\dag}
_{s'}\xi_{s}\right)-i\left(\vec{\epsilon}\,'^{*}\times\vec{\epsilon}\,\right)
\cdot\left(\xi'^{\dag}_{s'}\vec{\sigma}\,\xi_{s}\right)\right] \\
{\cal M}_{4} & = & \frac{-e^{2}}{2\omega' m^{2}}\left\{\sqrt{\sqrt{m^{2}+C^{2}}+
C}
\sqrt{\sqrt{m^{2}+B^{2}}+B}\left[(\omega'-2B)B+\omega'\sqrt{m^{2}+B^{2}}\right]
\right.\nonumber\\
& & +\left.\sqrt{\sqrt{m^{2}+C^{2}}-C}
\sqrt{\sqrt{m^{2}+B^{2}}-B}\left[(\omega'+2B)B-\omega'\sqrt{m^{2}+B^{2}}\right]
\right\} \nonumber\\
& & \times\left[\left(\vec{\epsilon}\cdot\hat{k}'\right)\vec{\epsilon}\,'^{*}
\cdot\left(\xi'^{\dag}_{s'}\vec{\sigma}\,\xi_{s}\right)+
i\left(\vec{\epsilon}\,'^{*}
\times\vec{\epsilon}\,\right)\cdot\hat{k}'\left(\xi'^{\dag}_{s'}\xi_{s}\right)
-\left(\vec{\epsilon}\,'^{*}\cdot\vec{\epsilon}\,\right)
\hat{k}'\cdot
\left(\xi'^{\dag}_{s'}\vec{\sigma}\,\xi_{s}\right)
\right].
\end{eqnarray}

An analysis of the general properties of the cross section is difficult.
The expression for ${\cal M}$ takes its simplest form when
the incoming and outgoing
photons are both polarized along the $y$-direction: $\vec{\epsilon}=\vec
{\epsilon}\,'=\hat{y}$. Then only the terms containing
$\vec{\epsilon}\,'^{*}\cdot\vec{\epsilon}$ are nonzero. Even with this
simplification, however, the
expression for the cross section
\begin{equation}
\label{eq-crosssection}
\frac{d\sigma}{d\Omega}=\frac{1}{64\pi^{2}}\frac{\omega'}{\omega\left(
\sqrt{m^{2}+B^{2}}-sB\right)\left[E'+\left(
1+\frac{s'B}{|\vec{p}\,'|}\right)(\omega'-\omega\cos\theta)\right]}
\left|{\cal M}_{1}+{\cal M}_{2}+{\cal M}_{3}+{\cal M}_{4}\right|^{2}
\end{equation}
is extremely unwieldy. Nor does extracting only the ${\cal O}(B)$ contribution
to $d\sigma/d\Omega$ simply things very greatly.
However, we would like to show that the cross
section (\ref{eq-crosssection}) has properties that distinguish it strongly from
the usual Klein-Nishina formula.
We shall therefore restrict our attention to a single
special limit---that of near-vanishing photon energy---in which
(\ref{eq-crosssection}) is completely dominated by the Lorentz-violating
contributions.

Specifically, we consider the case of $\omega\ll B^{2}/m\ll m$. Since $|B|$ is
expected to be small, this regime may not be accessible in a laboratory setting;
however, our results might still be testable astrophysically.
In this limit, we may set $C\approx B$ and $\omega\approx\omega'$
and neglect $B^{2}$ compared with $m^{2}$.
Then the matrix element becomes
\begin{equation}
{\cal M}\approx-i\frac{4e^{2}B^{2}}{\omega m}
\left(\vec{\epsilon}\,'^{*}\times\vec{\epsilon}\,\right)
\cdot\left(\xi'^{\dag}_{s'}\vec{\sigma}\,\xi_{s}\right).
\end{equation}
This is the dominant contribution
to ${\cal M}$ unless the cross product is near vanishing.
To determine a specific cross section, let us take $\vec{\epsilon}=\hat{y}$
and have $\vec{\epsilon}\,'=(\cos\theta,0,-\sin\theta)$ lie in the $xz$-plane.
Since the scattered electron has a nearly vanishing three-momentum, we may take
the quantization axis for the outgoing spin to be along the $z$-direction.
The phase space factor approaches its usual Lorentz-invariant form in this
limit, so the cross section is
\begin{equation}
\label{eq-specialcross}
\frac{d\sigma}{d\Omega}\approx\frac{e^{4}}{4\pi^{2}}\frac{B^{4}}{\omega^{2}
m^{4}}\left(\frac{1+ss'}{2}\cos^{2}\theta+\frac{1-ss'}{2}\sin^{2}\theta\right),
\end{equation}
which grows rapidly as $\omega\rightarrow0$. This is in sharp contrast
with the behavior of the Lorentz-invariant expression, which approaches the
frequency-independent Thomson result as $\omega\rightarrow0$. Although the
Thomson cross section is sometimes held to be a ``universal'' consequence of
gauge invariance~\cite{ref-thirring,ref-low,ref-gellmann,ref-abarbanel},
derivations of this
fact rely on additional assumptions, such as Lorentz symmetry, regularity of
the scattering amplitude at $\omega=0$, or the electron propagator having a
specialized form. Each of these assumptions is violated in this instance.

We see that the presence of the Lorentz violation can change the structure of
the Compton scattering cross section in a significant way. Our calculations
have been exact to all orders in $b$, and without this $b$-exact analysis, the
$\omega\rightarrow0$ cross section (\ref{eq-specialcross}) could not have been
determined correctly. In a perturbative calculation, the regime $\omega\ll
B^{2}/m$ would not have been accessible.
Moreover, in addition to determining the cross section for this particular
process,
we have also noted a general property of Lorentz-violating scattering;
when the Lorentz violation breaks the spin degeneracy of the energy-momentum
relations for the
external particles, Casimir's trick for performing polarization
sums may not work, because the velocity and phase space factors
in the cross section may depend upon the particles' spins.
Our results show the feasibility of scattering calculations for specific models
of Lorentz violation, and this work further demonstrates that such calculations
may even be performed
nonperturbatively. As such, this represents a major advance in the theory of
Lorentz-violating physics.

\section*{Acknowledgments}
The author is grateful to V. A. Kosteleck\'{y}, D. Colladay, and R. Jackiw for
their comments.
This work is supported in part by funds provided by the U. S.
Department of Energy (D.O.E.) under cooperative research agreement
DE-FG02-91ER40661.

\end{document}